

\def\sp{s^{\prime}}
\def\zp{z^{\prime}}
\def\pp{p^{\prime}}

\def\eq{$$}
\def\en{$$\medskip\noindent}
\def\sect{\bigskip\noindent}
\magnification=\magstep1
\hsize=6.0 true in
\hoffset=.1 true in
\vsize=8.5 true in
\voffset=.2 true in
\baselineskip=24 true pt
\nopagenumbers
\centerline{1/19/94}
\bigskip\bigskip

\line{\bf DIFFUSION-LIMITED AGGREGATION AS BRANCHED GROWTH  \hfill}
\bigskip\bigskip
\line{\hskip 1 true in \sl Thomas C. Halsey \hfill}
\line{\hskip 1 true in \sl The James Franck Institute and Department of Physics
\hfill}
\line{\hskip 1 true in \sl The University of Chicago \hfill}
\line{\hskip 1 true in \sl 5640 South Ellis Avenue \hfill}
\line{\hskip 1 true in \sl Chicago, Illinois 60637 \hfill}
\bigskip\bigskip\bigskip
\noindent{\bf Abstract}
\medskip

I present a first-principles theory of diffusion-limited
aggregation in two dimensions. A renormalized mean-field
approximation gives the form
of the unstable manifold for branch competition, following the method
of Halsey and Leibig [Phys.~Rev.~A {\bf 46}, 7793 (1992)]. This leads
to a result for the cluster dimensionality, $D \approx 1.66$, which is
close to numerically obtained values. In addition, the multifractal
exponent $\tau(3)=D$ in this theory, in agreement with a proposed
``electrostatic" scaling law. \vfill  \line{PACS Numbers: 64.60A,
68.70, 05.20 \hfill }\eject

\headline{\hfill \folio \hfill}
Diffusion-limited aggregation (DLA) is
a model of pattern formation in which clusters grow by the accretion
of successive random walkers.$^1$ Each random walker arrives from
infinity, and sticks to the growing cluster at whichever surface point
it first contacts. Only after the accretion of a walker does the next
walker commence its approach to the cluster. The clusters thereby
obtained are fractal in all dimensionalities $d>1$, and are
qualitatively and/or quantitatively similar to patterns observed in
such diverse phenomena as colloidal aggregation, electrodeposition,
viscous fingering, and dielectric breakdown.$^2$

At the heart of the problem of diffusion-limited aggregation is the
following question: what is the relationship between the
scale-invariance of the diffusive growth process and the hierarchical
structure of the clusters generated by this process?$^3$ A preliminary,
and incomplete, answer to this question was provided by this author
in collaboration with M. Leibig.$^4$ In this work, it was
hypothesized that the quantitative process by which one branch
screens, i.e., takes growth probability from, a neighboring branch,
has a specific form, independent of the length scale on
which this process takes place. This assumption allows the
development of a qualitatively correct theory, which yields
multifractal scaling of growth probability, as well as agreement with
a phenomonological scaling law, the ``Turkevich-Scher" law,
relating the scaling of the maximum growth probability over all sites
on the cluster to the dimension of the cluster as a whole.$^5$

In this letter, I shall present a more complete and {\it a priori}
theory of diffusion-limited aggregation in two dimensions based upon
a specific mean-field calculation of the dynamics of branch
competition. Because the mean-field approximation is implemented on all
length scales, it is perhaps better to regard this theory as an {\it
ansatz} solution in the case where certain types of fluctuations on
all length scales are neglected, while others are included.  This
specific model allows verification of all qualitative aspects of
branch competition that were advanced as (reasonable) hypotheses in
Ref.~4. The result obtained for the dimensionality of the cluster,
$D=1.66$, is within 3\% of the oft-quoted value $D=1.71$ obtained from
the scaling of the cluster radius-of-gyration in numerical studies.
An additional scaling law (the ``electrostatic scaling law"), relating
the multifractal exponent $\tau(3)$ of the growth measure to the
dimensionality $D$ by $D=\tau(3)$, is seen to be exact
within this theory.$^6$

In the growth process, each particle attaches itself to a unique
``parent" particle in the pre-existing cluster. Furthermore, the
cluster is observed to be a branched structure, with no loops and
with each particle having asymptotically zero, one or two
``children", i.e. particles to whom it stands as a parent.$^7$  Very
rarely particles have more than two children; primarily for
reasons of convenience I neglect this possibility.

Consider a
particle with two children. Each of the two children separately, with
all particles descended from each, I term a ``branch". Thus these
two-child particles are parents of two branches, which occupy
neighboring regions of space. The total number of particles in one
branch I term $n_1$, and the total in the other $n_2$. The total
number of descendants of the parent particle is thus $n_b \equiv n_1
+n_2$. Now consider the next particle to accrete to the cluster. I say
that this particle has a total probability $p_1$ to stick anywhere on
the first branch, and a total probability $p_2$ to stick anywhere on
the second branch, yielding a total probability $p_b \equiv p_1 +
p_2$.

Let us now consider the normalized quantities $x=p_1/p_b$ and
$y=n_1/n_b$. Clearly $dn_1 / dn = p_1$, where $n$ is the total number
of particles in the cluster, and we are neglecting fluctuations of
$O(\sqrt n_b)$. Thus $y$ obeys the following equation of motion:

\eq
{d y \over d \ln n_b} = x - y .
\eqno(1)
\en
The right-hand side of this equation is a function only of $x$ and
$y$. Now $x$ will obey an equation of the form

\eq
{d x \over d \ln n_b} = G(x,y; n; \{\phi_i\}) ,
\eqno(2)
\en
where $\{ \phi_i \}$ is some parameterization of all of the variables
describing the structure of the cluster. In ref.~4 we assumed that by
averaging the right-hand side of this equation over these parameters $\{
\phi_i \}$, one obtains $d x /d \ln n_b =g(x,y)$, where the right-hand
side is now only a function of $x$ and $y$. Given this function
$g(x,y)$, one has a closed system of equations describing the
evolution of $x$ and $y$ as functions of $\ln n_b$.

By symmetry, $g(x,y) = -g(1-x,1-y)$, so $(x,y)=(1/2,1/2)$ must be a
fixed point of this process of competition between the two branches.
In ref.~4, we explored the consequences of assuming that this fixed
point is hyperbolic, with the unstable manifold emerging from the
fixed point terminating in two stable fixed points at $(x,y)=(0,0)$
or $(x,y)=(1,1)$, these latter representing the situation in which
one branch has been completely screened by the other. This assumption
will be explicitly verified in the calculation below.

If the central fixed point at $(x,y)=(1/2,1/2)$ is hyperbolic, then
branch pairs which commence their existence (with $n_b \sim 1$) near
the unstable fixed point will be quickly drawn onto the unstable
manifold. Linearizing the system of equations for $d (x,y) / d \ln
n_b$ about the central fixed point, the hyperbolic assumption implies
that there will be a stable and an unstable direction; the
eigenvalue corresponding to the latter direction we
define to be $\nu$.

When a pair of branches is first created by a tip-splitting event,
its initial growth up to the stage at which $n_b \gg 1$ is
determined by complicated microscopic dynamics, which do not recognize
the existence of the unstable fixed point. Thus we expect the
probability that a newly created branch pair will be a distance
$\epsilon^{\nu}$ from the unstable fixed point will be $\rho (
\epsilon \ll 1) d \epsilon \propto \epsilon^{\nu-1} d \epsilon$; we
are assuming a constant probability density of branch creation near
the unstable fixed point. This assumption has been specifically
verified by numerical study in ref. 4.
The choice of $\epsilon^{\nu}$ for this initial distance insures that
position along the unstable manifold in the $x-y$ plane can be
parameterized by the variable $\epsilon n_b$.

It is possible to relate the eigenvalue $\nu$ to the cluster
dimensionality $D$ by the following argument.$^4$  Consider the
strongest branch in the cluster, that obtained by always following the
stronger child (with the larger values of $x$,$y$) at each branching.
The total number of side-branches (or branch points) from such a
branch is $\sim r$, where $r$ is the cluster radius. In order that the
cluster have a dimension $D >1$, a number $\sim 1$ of these side
branches must have a total number of particles $\sim n$, the total
number in the cluster. A side branch obeying this criterion must have
$\epsilon n \sim 1$, so that at that branching, both descendant
branches are roughly equal in size. The probability of this happening
at any particular branching is $\int^{n^{-1}} d \epsilon
\; \rho(\epsilon) \propto n^{\nu}$, and there are $\sim r$ different
sidebranchings at which this might occur. Thus $r n^{\nu} \sim 1$, or
$D = 1 / \nu$.

In order to determine $g(x,y)$, we turn to an explicit description of
the growth process.$^{6,8}$ Suppose that we parameterize the accessible
surface of the cluster by arc-length $s$. If a
particle attaches at the surface point $\sp$,
it thereby reduces the  growth probability at all points $s$ for
which $\vert s - \sp \vert > a$, where $a$ is the particle size. This
is because a certain number of the random walks that would have reached
$s$ previously are now obstructed by the new particle at $\sp$. If the
probability that a particle lands at $\sp$ is $p( \sp )$, and the
probability that a random walker goes from $\sp$ to $s$ without
contacting the surface is $H(s,\sp)$, this implies
that

\eq
{dp(s) \over d n} = \int d \sp \left ( H(s,\sp)-h(s)
\delta(s-\sp) \right ) p^2(\sp)  ,
\eqno(3)
\en
where we have modelled effects on the scale $\vert s-
\sp \vert < a$ by the $\delta$-function, the
coefficient of which, $h(s)$, is set by the conservation
of the total growth probability, $\int ds p(s) = 1$. Note that in
Eq.~(3), two factors of $p(\sp)$ appear--one corresponds to the
original probability that a particle lands at $p(\sp )$, the other to
the potential trajectories arriving at $s$ that are blocked by such a
particle.

For $a \ll \vert s-
\sp \vert \ll a n$, conformal transformation shows that the function
$H(s,\sp)$ is given in two dimensions by the simple form$^9$

\eq
H(s,\sp) = {p(s) p(\sp) \over \left [ \int_s^{\sp} ds^{\prime\prime}
p(s^{\prime\prime}) \right ]^2}  ,
\eqno(4)
\en
where the integral in the denominator is the
total growth probability between the points $s$ and
$\sp$. It is
convenient to parameterize the interface by this quantity, the ``growth
probability" distance between points $z(s)$, defined by $z(\sp) - z(s)
= \int_s^{\sp} ds^{\prime\prime}  \; p(s^{\prime\prime})$. Then our
fundamental equation becomes

\eq
{d p(z) \over d n} = p(z) \int d \zp \left [ {1 \over (z - \zp )^2} -
\tilde h(z) \delta(z-\zp) \right ] p^2(\zp)  ,
\eqno(5)
\en
where $a$ serves as an ultra-violet cutoff to prevent divergence of
the integral, and $\tilde h(z)$ is related to $h(s)$ and to the
function $z(s)$; its precise form is of no interest to us.

I wish to use this equation to determine the function $ d x /
d \ln n_b = g(x,y)$. Repeated application of the chain rule yields

\eq
{ d x \over d \ln n_b} = {n_b \over  p_b^2} \left \{ (1-x)
{dp_1 \over dn} - x {dp_2
\over dn} \right \}  .
\eqno(6)
\en
Consider a branch with probability $\pp$ and a number of
particles $n^{\prime}$. We suppose that this branch extends from
$z=0$ to $z=\pp$. Eqs.~(5) and (6) imply that if we can
write $p^2 (z)$ on this branch (and by extension, all other branches)
as

\eq
p^2 (z) =  {(\pp)^2 \over n^{\prime} } f(z / \pp )  ,
\eqno(7)
\en
where $f(z)$ is a universal function that depends neither upon
$\pp$ nor upon $n^{\prime}$, then we will be able to write $dx
/ d \ln n_b = g(x,y)$, with the right-hand side a function of $x$ and
$y$ alone. Equation~(7) is motivated by the fact that $p^2 (z)$ must
be proportional to $(\pp)^2$; the dependence on $n^{\prime}$ is
specifically chosen to lead to an $n^{\prime}$-independent $g(x,y)$.
Only if we can find a method of computing an $n^{\prime}$-independent
$f(z)$ will this {\it ansatz} be justified.

Thus the crux of the problem is this ``branch envelope" function
$f(z)$, which represents, with the appropriate normalization, the
distribution of growth probability in different regions of a branch.
Now in our picture, each branch can be divided into two distinct
sub-branches, which compete according to the dynamics established by
$g(x,y)$. Our central mean-field assumption is that we can compute
$f(z)$ by averaging the envelope functions $f(z)$ of these
sub-branches over the stochastic parameter $\epsilon$ appropriate to
the competition of these two sub-branches. In this way we obtain the
following equation:

\eq
f(z) = \int_{-\infty}^{\infty} d \epsilon \rho (\epsilon )\left \{
{x^2 (\epsilon n_b ) \over y(\epsilon n_b)} f \left ( {z \over x(
\epsilon n_b ) } \right ) + {(1-x(\epsilon n_b ) )^2 \over
(1-y(\epsilon n_b ) ) }f \left ( {1-z \over 1-x( \epsilon n_b ) }
\right ) \right \} , \eqno(8)
\en
where $x(\epsilon n_b)$ and $y(\epsilon n_b)$ give the values of $x$
and $y$ along the unstable manifold as functions of $n_b$ and the
stochastic parameter $\epsilon$. For convenience, we are defining
$\rho (\epsilon)$ for negative values of $\epsilon$ as $\rho(-
\epsilon) = \rho (\epsilon)$, with $x(-\eta) = 1-x(\eta)$, $y(-\eta)
= 1 - y(\eta)$. This leads to the relatively compact expression of
Eq.~(8). For large $n_b$, this equation has a solution independent of
$n_b$, which is determined by

\eq
\int_{-\infty}^{\infty} d \eta \vert \eta \vert^{\nu-1} \left \{ {x^2
(\eta ) \over y(\eta) } f\left ({z \over x(\eta) } \right ) +
{(1-x(\eta ) )^2\over (1-y(\eta ) )} f \left ( {1-z \over 1-x(\eta) }
\right )  - f(z) \right \} = 0  .
\eqno(9)
\en
Since the integrand goes to zero as $\eta \to \infty$, we are
justified in taking the small $\epsilon$ form for $\rho (\epsilon )$.

Of course, in order to perform this integral, we must have the form
of the unstable manifold, and thus we must already know $g(x,y)$. We
can determine $g(x,y)$ from $f(z)$ by simply integrating Eq.~(5) over
the appropriate intervals. We do not integrate over regions
exterior to the two competing branches, but only investigate
the influence of the two branches on one another. Skipping some tedious
algebra, we may express the result as follows. Defining a function
$\psi(u)$ by

\eq
\psi(u) =  \int_0^1 dz \left ( {1 \over z} - {1 \over z +
u} \right ) f(z)  ,
\eqno(10)
\en
we can write

\eq\eqalign{
g(x,y) = & x (1-x)  \Bigg \{ \left [ 2 {x \over y} \psi(\infty) -
{(1-x)^2 \over (1-y) x} \psi \left ( {x \over 1-x} \right ) \right ]
\cr & - \left [
2{1-x \over 1-y} \psi(\infty)  - { x^2 \over y(1-x) }
\psi \left ( 1-x \over x  \right ) \right ] \Bigg \}. \cr }
\eqno(11)
\en

The reader should note that we have a circular procedure,
because $g(x,y)$ is determined as a function of $f(z)$ by Eqs.~(10)
and (11), while $f(z)$ is determined as a function of $g(x,y)$, and in
particular by the unstable manifold in the $x-y$ plane as
determined by $g(x,y)$, by Eq.~(9). Thus in practice we are looking for
a solution of Eq.~(9) where the functions $x(\eta)$ and $y(\eta)$ are
implicitly determined by $f(z)$.

I have numerically obtained the unique solution to Eq.~(9)
under these conditions, which is displayed in the inset to Figure
1.$^{10}$ This validates our assumption regarding the scaling with
$n^{\prime}$ in Eq.~(7). The function $g(x,y)$ determined from this
function has all of the necessary qualitative features; in particular,
the fixed point at $(x,y) = (1/2,1/2)$ is unstable and hyperbolic, and
the unstable manifold leads from this point to stable fixed points at
$(x,y) = (0,0)$ and $(1,1)$, as illustrated in Figure 1. Figure 1 also
shows numerical results for branch competition. The value of the
unstable eigenvalue $\nu$ is $\nu \approx .6020$, implying that
$D=1/\nu \approx 1.661$, which is within 3\% of the standard numerical
result $D \approx 1.71$.

In addition, this theory automatically agrees with the electrostatic
scaling law, which states that

\eq
\langle \int ds \; p(s)^3 \rangle \propto n^{-1}  ,
\eqno(12)
\en
where the integral is over the entire cluster surface. This is
equivalent to the more usual statement that $\tau(3) = D$. In ref.~4,
we demonstrated that the multifractal exponents $\sigma(q)$ defined by
$\int ds \; p(s)^q \propto n^{-\sigma(q)}$ can be obtained from the
integral condition$^{11}$

\eq
\int_0^{\infty} d \eta \; \eta^{\nu-1} \left \{ {x(\eta)^q
\over y(\eta)^{\sigma(q)} } + {(1-x(\eta))^q \over
(1-y(\eta))^{\sigma(q)} } -1 \right \} = 0  .
\eqno(13)
\en
By integrating Eq.~(9) from $z=0$ to $z=1$, one obtains
precisely this criterion, with $q=3$ and $\sigma(q)=1$, in agreement
with the electrostatic scaling law.  Though the electrostatic scaling
law thus appears in a natural way in this theory, one should not say
that it is predicted by this theory unless the solution obtained to
Eq.~(9) is stable. It may be that it is necessary to impose the
electrostatic scaling law as a constraint to insure this
stability.$^{10}$

{}From Figure 1, it is clear that although in some sense the unstable
manifold that we have calculated is an acceptable average
trajectory, the numerically obtained trajectories do exhibit some
dispersion about this average. This has significant results. The
Makarov scaling law predicts that $d \sigma (q) /dq \vert_{q=1} =
1/D_0$,$^{12}$ where $D_0$ is the surface fractal dimension (which
according to some studies is significantly less than the
radius-of-gyration dimension $D$.)$^{13}$ My result, from Eq.~(13), is
$d \sigma (q) /dq \vert_{q=1} \approx 0.71$, which is significantly
different from the Makarov result. In practice, this quantity is
quite sensitive to the way in which the unstable manifold approaches
the stable fixed points at $(x,y) = (0,0)$ and $(1,1)$; since the
numerical trajectories are quite dispersed in this region, I do not
expect a good result for the Makarov scaling from a one-trajectory
theory. However, the theory outlined in this letter can be easily
generalized to account for the possibilty of trajectory dispersion,
which may lead to better agreement with the Makarov result.

\sect{\bf Acknowledgements} \medskip

I would like to acknowledge a stimulating discussion with
L.P. Kadanoff, as well as conversations with R. Blumenfeld on a
closely related topic. I am very grateful to A. Libchaber for
encouragement at an early stage in this project. This work was
supported by the National Science Foundation through a Presidential
Young Investigator award, Grant DMR-9057156.

\vfill\eject
\sect{\bf Notes}
\medskip
\item{1.} T.A. Witten, Jr. and L.M. Sander, Phys.~Rev.~Lett.~{\bf
47}, 1400 (1981); P. Meakin, Phys.~Rev.~A {\bf 27}, 1495 (1983).

\item{2.} R. Brady and R.C. Ball, Nature (London) {\bf 309}, 225
(1984); L. Niemeyer, L. Pietronero, and H.J. Wiesmann,
Phys.~Rev.~Lett.~{\bf 52}, 1033 (1984); J. Nittmann, G. Daccord, and
H.E. Stanley, Nature (London) {\bf 314}, 141 (1985).

\item{3.} This question is also the focus of real-space studies such
as L. Pietronero, A. Erzan, and C. Evertsz, Phys.~Rev.~Lett. {\bf 61},
861 (1988); Physica A {\bf 151}, 207 (1988), and X.R. Wang, Y. Shapir
and M. Rubenstein,  Phys.~Rev.~A {\bf 39}, 5974 (1989); J.~Phys.~A
{\bf 22}, L507 (1989).

\item{4.} T.C. Halsey and M. Leibig, Phys.~Rev.~A {\bf 46}, 7793
(1992); T.C. Halsey and K. Honda, unpublished.

\item{5.}  L. Turkevich and H. Scher, Phys. Rev. Lett. {\bf 55}, 1026 (1985);
Phys. Rev. A {\bf 33},
786 (1986); see also  R. Ball, R. Brady, G. Rossi, and
B.R. Thompson, Phys.~Rev.~Lett. {\bf 55}, 1406 (1985), and T.C. Halsey,
P. Meakin, and I. Procaccia, Phys.~Rev.~Lett. {\bf 56}, 854 (1986).

\item{6.} T.C. Halsey, Phys.~Rev.~Lett. {\bf 59}, 2067 (1987);
Phys.~Rev.~A {\bf 38}, 4749 (1988).

\item{7.} The fact that there are no loops follows from the fact
that every particle has a unique parent, which is true in
off-lattice versions of DLA.

\item{8.} B. Shraiman and D. Bensimon, Phys.~Rev.~A {\bf 30}, 2840
(1984); R.C. Ball and M. Blunt, Phys.~Rev.~A {\bf 39}, 3591 (1989).

\item{9.} T.C. Halsey, Phys.~Rev.~A {\bf 35}, 3512 (1987).

\item{10.} The stability of this solution is a more difficult
question. There appears numerically to be a single instability of the
solution, which can be eliminated if one applies the electrostatic
scaling law as a constraint.

\item{11.} For a general discussion of multifractality, see T.
Vicsek, {\it Fractal Growth Phenomena, 2nd ed.} (World Scientific,
Singapore, 1992).

\item{12.} N.G. Makarov, Proc.~London Math.~Soc. {\bf 51}, 369 (1985).

\item{13.} In particular, $D_0 \approx 1.60$ is claimed by F. Argoul,
A. Arneodo, G. Grasseau, and H. Swinney, Phys.~Rev.~Lett.~{\bf 61},
2558 (1988).

\vfill \eject
\sect{\bf Figure Caption}
\medskip
\headline{\hfill}

\item{1.} Trajectories of branch competion in the $x-y$ plane. The
light solid trajectories are numerical results from ref.~4 for
specific branch pairs in growing DLA clusters. The heavy solid line
represents the unstable manifold predicted by this letter, which is
quite close to the ``average" numerical trajectory. The
inset shows the computed branch envelope function $f(z)$.

\vfill\eject\end